# Integration of Quantum Emitters with Lithium Niobate Photonics

*Shahriar Aghaeimeibodi[1], Boris Desiatov[2], Je-Hyung Kim[3], Chang-Min Lee[1], Mustafa Atabey Buyukkaya[1], Aziz Karasahin[1], Christopher J. K. Richardson[4], Richard P. Leavitt[4], Marko Lončar[2], and Edo Waks[1,5,a]*

[1]Department of Electrical and Computer Engineering and Institute for Research in Electronics and Applied Physics, University of Maryland, College Park, Maryland 20742, United States

[2]School of Engineering and Applied Sciences, Harvard University, Cambridge, Massachusetts 02138, USA

[3]Department of Physics, Ulsan National Institute of Science and Technology, Ulsan 44919, South Korea

[4]Laboratory for Physical Sciences, University of Maryland, College Park, Maryland 20740, United States

[5]Joint Quantum Institute, University of Maryland and the National Institute of Standards and Technology, College Park, Maryland 20742, United States

---

[a] Corresponding Author, E-mail: edowaks@umd.edu

ABSTRACT. The integration of quantum emitters with integrated photonics enables complex quantum photonic circuits that are necessary for photonic implementation of quantum simulators, computers, and networks. Thin-film lithium niobate is an ideal material substrate for quantum photonics because it can tightly confine light in small waveguides and has a strong electro-optic effect that can switch and modulate single photons at low power and high speed. However, lithium niobite lacks efficient single-photon emitters, which are essential for scalable quantum photonic circuits. We demonstrate deterministic coupling of single-photon emitters with a lithium niobate photonic chip. The emitters are composed of InAs quantum dots embedded in an InP nanobeam, which we transfer to a lithium niobate waveguide with nanoscale accuracy using a pick-and place approach. An adiabatic taper transfers single photons emitted into the nanobeam to the lithium niobate waveguide with high efficiency. We verify the single photon nature of the emission using photon correlation measurements performed with an on-chip beamsplitter. Our results demonstrate an important step toward fast, reconfigurable quantum photonic circuits for quantum information processing.

Thin-film lithium niobate (LiNbO$_3$) is an emerging material platform for integrated photonics[1–3] that exhibits tightly confined optical modes, high refractive index, and wide transmission window (350 nm to 5 micron). Moreover, this material inherits the strong $\chi^2$ electro-optic nonlinearity from LiNbO$_3$ bulk crystals,[4–6] which enables ultrafast optical modulation. The tight confinement of the optical modes in thin-film LiNbO$_3$ significantly reduces the device size and facilitates scalable fabrication of many optical elements on a small chip. This scalability was not possible using conventional metal diffused waveguides, as they have large and loosely confined optical modes.[7] Recently, researchers have realized nanophotonic structures based on thin-film LiNbO$_3$ with low loss[8,9] and high modulation bandwidth[10] competing with the metal diffusion technology, but with a much smaller footprint, turning thin-film LiNbO$_3$ into a versatile platform for integrated photonic circuits.

Strong electro-optic non linearity and compact nature of thin-film LiNbO$_3$ make it an ideal platform for quantum photonic circuits that can enable optical quantum computation,[11–16] high-speed quantum communications,[17,18] and simulation of non-classical problems in quantum physics,[19] chemistry,[20] and biology.[21] Many of these applications require quantum emitters that serve as both high-purity sources of indistinguishable single photons,[22,23] and strong optical nonlinearities at the single photon level.[24,25] As a result, developing techniques for the integration of LiNbO$_3$ photonics with quantum emitters is an effective strategy for implementing fast reconfigurable quantum circuits. But to date this integration has yet to be demonstrated. One of the difficulties is that conventional metal diffused LiNbO$_3$ waveguides exhibit a small index contrast and large mode volume, which leads to poor transfer efficiencies for emitters that are embedded or evanescently coupled. But thin-film LiNbO$_3$ exhibits a much tighter mode confinement that could potentially solve this problem, which provides a new opportunity for quantum emitter integration.

In this letter, we demonstrate integration of quantum emitters with LiNbO$_3$ photonic devices. The quantum emitters are InAs quantum dots embedded in an InP nanobeam, which serve as efficient sources of single photons in the telecom band.[26,27] We develop a hybrid device structure that efficiently transfers the emission from the dots to a LiNbO$_3$ waveguide. The tight mode confinement of the LiNbO$_3$ waveguide enables efficient transfer of photons from the InP nanobeam to the waveguide through evanescent coupling with efficiency exceeding 34%, which would be extremely difficult to achieve with larger metal diffused waveguides. To experimentally demonstrate this approach, we fabricate a hybrid device using a pick-and-place technique based on focused ion beam.[28] We verify efficient transfer of single photons from the quantum dot to the LiNbO$_3$ and confirm the single-photon nature of the emission with photon correlation measurements. This approach could enable scalable integration of single-photon emitters with complex LiNbO$_3$ photonic circuits that can rapidly modulate the photons and perform user-defined linear optical transformations on them.

Figure 1a and 1b show the general scheme for coupling single-photon emitters with a LiNbO$_3$ waveguide. Figure 1a shows a cross-sectional illustration of the device which is composed of an InP nanobeam (500 nm wide and 280 nm thick) containing InAs quantum dots on top of a LiNbO$_3$ waveguide. In the design we use a partially etched LiNbO$_3$ waveguides with 1200 nm width, which ensures the single mode condition at the InAs quantum dot wavelength of ~1300 nm while maintaining a relatively large top surface area to transfer the InP nanobeams. The single photon from the quantum dot couples to the InP nanobeam and then smoothly transfers to the LiNbO3 waveguide through a 5 μm adiabatic taper (Figure 1b). A Bragg reflector at one end of the nanobeam, composed of a periodic array of holes with a period of 290 nm and radii of 100 nm, ensures the quantum dot emission propagates in only one direction (Figure S1).

We performed finite difference time domain (Lumerical) simulations to estimate the efficiency of single photon coupling from the quantum dots to the InP nanobeam and subsequently to the $LiNbO_3$ waveguide. In our simulation, we model the quantum dots as electric dipole emitters with an in-plane polarization that are located at the center of the nanobeam. Figure 1c displays a cross sectional view of light propagation in the hybrid device. The simulation shows that emission from the quantum dot couples to the single mode of the InP nanobeam, and then adiabatically transfers to the $LiNbO_3$ waveguide as the taper narrows down. We calculate the coupling efficiency between the InP nanobeam and $LiNbO_3$ waveguide modes for a taper length of 5 μm to be 40.1%. A longer adiabatic taper can further improve this efficiency (See Figure S1c). However, for our current devices, we used a 5 μm taper length to make it easier to transfer the nanobeam onto the waveguide using the pick-and-place method described below. The total efficiency from the quantum dot to the $LiNbO_3$ waveguide mode was calculated to be 34% by multiplying the efficiency of coupling for the quantum dot to the InP mode (85%) and the efficiency of InP to $LiNbO_3$ coupling (40.1%).

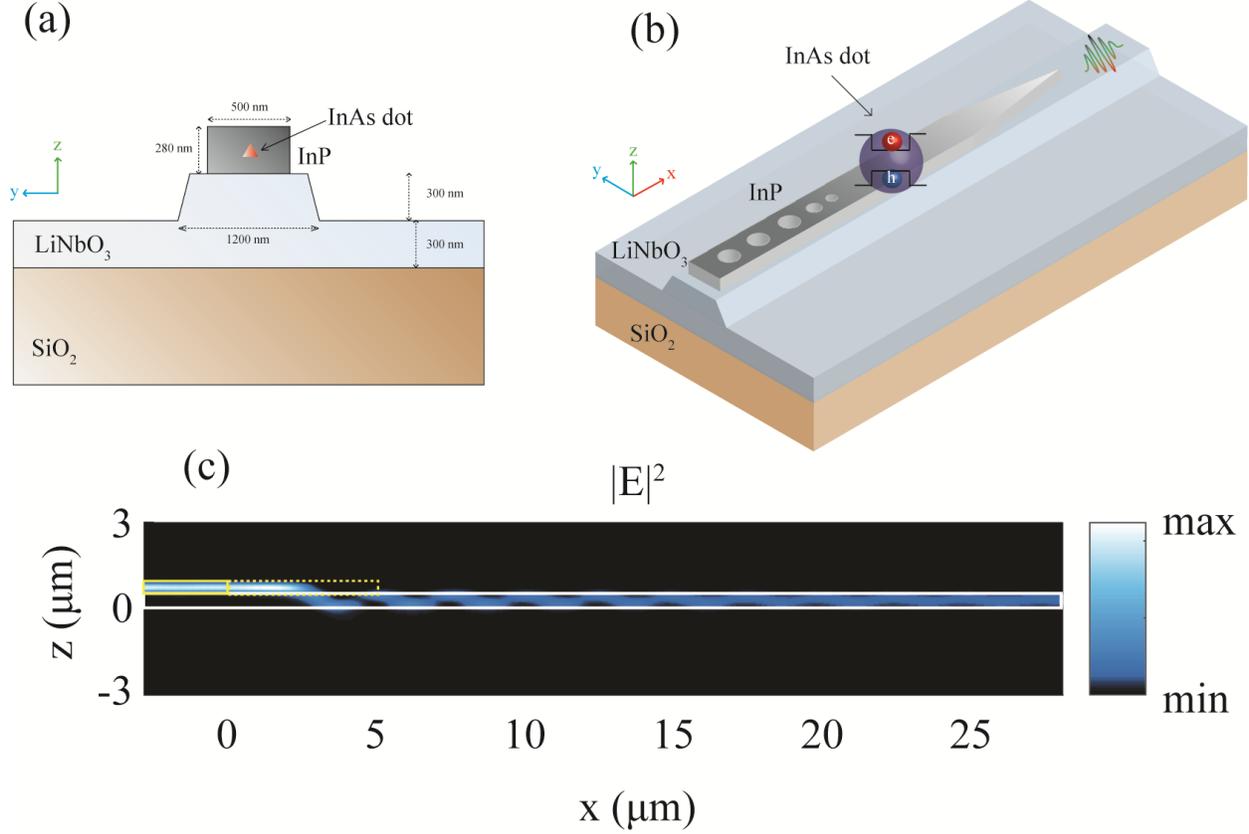

**Figure 1.** (a) An illustration of the hybrid platform consisting of the LiNbO$_3$ waveguide and the InAs quantum dot embedded in InP nanobeam. (b) A side schematic of the hybrid device, showing the Bragg reflector on one side of the InP nanobeam and the adiabatic taper on the other side. (c) Finite difference time domain simulation of the electric field intensity $|E|^2$ in the coupling area. Yellow and white solid lines represent the boundaries of the InP nanobeam and LiNbO$_3$ waveguide, respectively. The dashed yellow lines indicate the boundary of the tapered region of the nanobeam.

To fabricate the designed device, we first patterned the nanobeam and LiNbO$_3$ waveguide on separate substrates, and then transferred the nanobeam to the waveguide using a pick-and-place technique we previously developed.[28] The substrate for the LiNbO$_3$ waveguides was a 600 nm thick X-cut LiNbO$_3$ film on 2 μm thick silicon dioxide (SiO$_2$) and a silicon substrates (NanoLN).

We patterned the photonic structures with electron beam lithography using a HSQ resist. Next, we transferred the patterns onto a LiNbO$_3$ thin film using an optimized Ar+ plasma etching recipe in a reactive ion etching tool. Finally, we removed the residual mask by buffered oxide etching. Figure 2a shows the fabricated waveguide structure as well as y-branch 50:50 beamsplitter. We terminated the waveguides with a periodic grating coupler with a period of 700 nm at one end of the structure for outcoupling the single-photon emission, with a calculated efficiency of 26.7% (see figure S2 in supplementary material for the design of the gratings).

To create the InP waveguides, we began with a substrate composed of 280 nm InP on a 2 μm thick AlInAs sacrificial layer. We patterned the InP membrane with electron beam lithography, followed by dry etching and selective wet etching of the sacrificial layer to form a suspended structure. Figure 2b displays a scanning electron microscopy (SEM) image of the suspended InP nanobeam. The square pad at one end of the nanobeam facilitates the pick-and-place procedure. By contacting the pad with a microprobe tip and cutting the remaining InP tethers with the focused ion beam, we release the nanobeam from the substrate (Figure 2c) and place it on a previously fabricated LiNbO$_3$ straight waveguide (Figure 2d) or beamsplitter (Figure 2e).

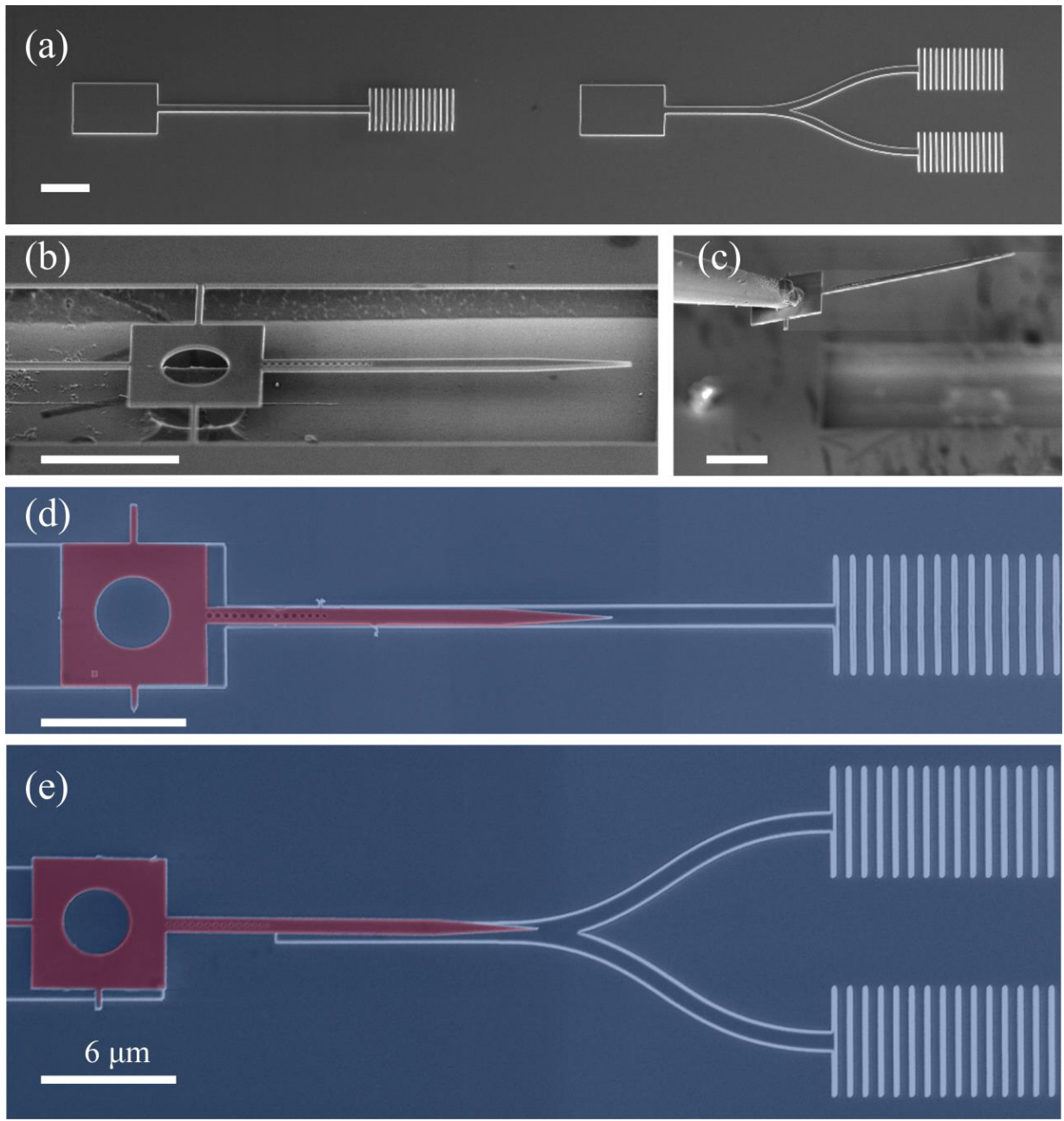

**Figure 2.** (a) SEM image of the LiNO₃ straight waveguide and y-branch beamsplitter. The rectangular pads on the left end of the LiNbO₃ waveguides assist in the transfer process. Periodic grating couplers on the right side of the devices extract the light from the LiNbO₃ waveguides. (b) SEM image of the suspended InP nanobeam. (c) The InP nanobeam attaches to the microprobe tip by van der Waals forces. (d) False color SEM image of the integrated InP nanobeam with a straight LiNbO₃ waveguide. (e) False color SEM image of the integrated InP

nanobeam with an on-chip LiNbO$_3$ beamsplitter. Red and blue colors indicate InP and LiNbO$_3$, respectively. Scale bar is 6 μm in all panels.

To characterize the fabricated devices, we performed optical measurements using a low-temperature micro-photoluminescence setup operated at 4 K (see supplementary material). We excited the quantum dots with a 780 nm continuous wave laser and collected the photoluminescence spectrum through the grating coupler (inset of Figure 3a). Figure 3a shows the photoluminescence spectrum taken from the straight LiNbO$_3$ waveguide. We observed multiple emission lines in the photoluminescence spectrum that confirmed the coupling of multiple spectrally resolved quantum dots. To assess the photon collection efficiency in our hybrid device, we used a 785 nm pulsed laser excitation with 40 Mhz repetition rate. Correcting for our setup efficiency, we calculated a collection efficiency of 2.2% at the first lens for a representative coupled quantum dot labeled as QD1 in Figure 3a (see supplementary material). This value is lower than the ideal collection efficiency of 9% that we determined from our simulations. The simulated collection efficiency is the product of the coupling efficiency from the quantum dot to LiNbO$_3$ (34%) and the grating coupler efficiency (26.7%). We atttribute the lower experimental collection efficiency to small fabrication imperfections (Figure S2), residual misalignment of the nanobeam with the LiNbO$_3$ waveguide, and quantum dot deviations from the center of the nanobeam.

To confirm the single photon nature of the emission, we performed second order photon correlation measurements on several of the coupled quantum dot emission lines. In this setup, we sent the collected signal through the grating coupler to a fiber beamsplitter and connected the two output ports of the beamsplitter to different single-photon detectors. Figure 3b shows a continuous wave second-order correlation measurement for QD1, using a 780 nm laser. The measurement shows a clear antibunching behavior. We fit the antibunching dip to a function of the form $g^{(2)}(\tau) = 1 - (1 - g^{(2)}(0)) \exp(-|\tau|/\tau_0)$ without dark count subtraction or deconvolution and obtained $g^{(2)}(0) = 0.08$, which is lower than the classical limit of 0.5. Background emissions due to our non-resonant excitation cause the residual multiphoton events resulting in non-ideal $g^{(2)}(0)$ values in our photon correlation measurements. Resonant excitation[29,30] or quasi-resonant excitation[31,32] could significantly improve the purity of the single photons by reducing the background emissions.

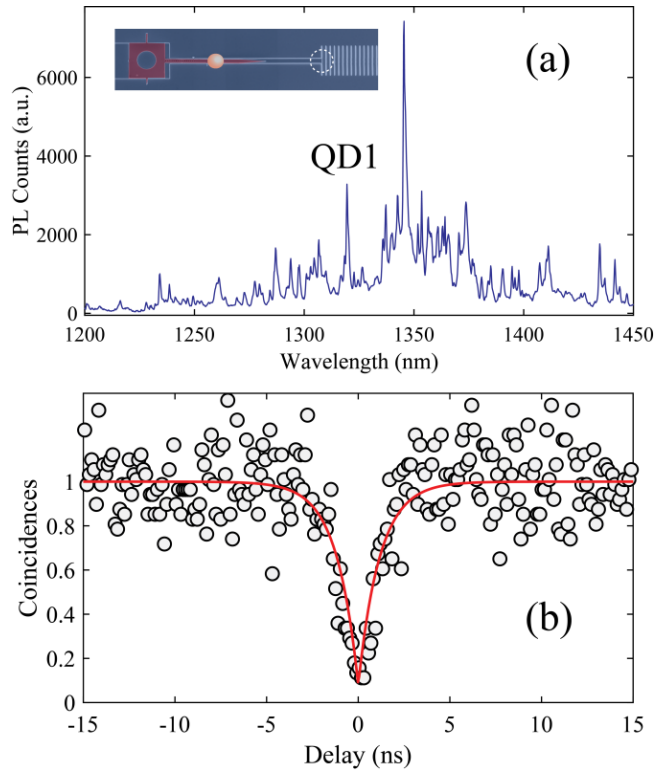

**Figure 3.** (a) Photoluminescence spectrum of the LiNbO$_3$ coupled quantum dots when we excite the quantum dots on top of the nanobeam and collect the signal through the grating. The inset indicates the excitation and collection scheme, in which the orange dot and the white dashed circle represent the excitation and collection spots, respectively. (b) Second order photon correlation measurement of QD1 when excited with a continuous wave laser.

Next, we investigated the hybrid device in Figure 2e that integrates quantum dots in the InP nanobeam with an on-chip LiNbO$_3$ beamsplitter. We excited the quantum dots directly from the top of the nanobeam and collected the photoluminescence signal from both grating couplers (Insets of Figure 4a and 4b). We separated the signal from each grating using a pick-off mirror in free space and sent each port to separate spectrometers that acted as spectral filters. Figures 4a and 4b show the collected photoluminescence signal from the top and bottom gratings. We observed multiple quantum dot lines in both spectra. We identified 7 emission lines that appear in both spectra, suggesting that they originate from the same quantum dots.

To confirm that the replicated emission lines in figure 4a and 4b originate from the same quantum dot, we performed a photon correlation measurement on the quantum dot represented by line 4, and spectrally filter out all other emission lines (See Figure S3 for a schematic of the measurement setup). Figure 4c shows the second-order photon correlation measurement of this emission using continuous wave excitation, with $g^{(2)}(0) = 0.36$. The $g^{(2)}(0)$ value obtained for this quantum dot is higher than the one for QD1 in figure 3b, which could be because of difference in intrinsic single-photon purity between these two emitters. This measurement demonstrates that the two matched lines from the grating couplers originate from the same quantum dot. The on-chip LiNbO$_3$ beamsplitter enables the direct measurement of a second-order photon correlation from the quantum dots coupled to the LiNbO$_3$ waveguide without an external beamsplitter. Implementing this functionality on-chip is a step toward scalable integration of multiple single-photon emitters with more complex LiNbO$_3$ photonic circuits, where most of the light manipulations happen on a compact photonic chip.

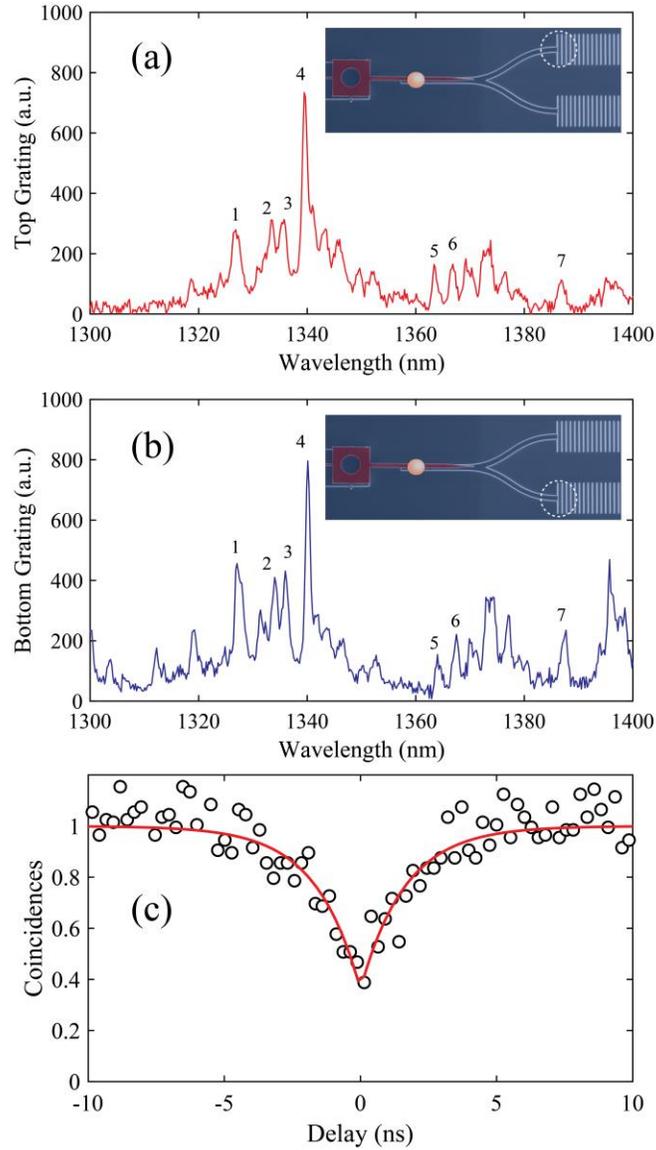

**Figure 4.** (a,b) Photoluminescence spectrum collected through the (a) top and (b) bottom gratings. Insets indicate the excitation and collection scheme. (c) Second order correlation measurement of emission line 4, labeled in (a,b).

In summary, we have deterministically coupled quantum emitters to a nanophotonic $LiNbO_3$ waveguide. We confirmed the coupling of single photons to straight waveguides and y-branch beamsplitters using photon correlation measurements. The collection efficiency of our devices was limited by the performance of the grating couplers. More efficient grating coupler design or better approaches of light collection, such as edge coupling,[33] tapered fibers,[34] or the incorporation of detectors on the chip,[35] could boost the collection efficiency. Incorporation of Mach-Zehnder or resonator based electro-optic switches[10] would allow fast switching of single photons. Moreover, by pre-characterizing the quantum dot devices before pick-and-place, we can overcome the spectral and spatial randomness of the quantum dots and select the devices with emitters at the same resonance wavelength. In this way, we can efficiently extend our results to complex $LiNbO_3$ circuits containing many identical single-photon emitters. Our results represent an important step toward routing and fast feedforwarding of on-demand single photons on a chip, which could enable high speed quantum communication[17,18] and linear optical approaches for photonic quantum computing.[14,16]

SUPPLEMENTARY MATERIAL

Additional finite-difference time-domain simulation is provided for the taper coupling efficiency, Bragg mirror geometry, and design and simulation of the grating couplers. Also this file includes details on measurements setup and estimation of the quantum dot collection efficiency.

ACKNOWLEDGMENT

The authors would like to acknowledge support from the Laboratory for Telecommunication Sciences, The Center for Distributed Quantum Information at the University of Maryland and Army Research Laboratory, and the Physics Frontier Center at the Joint Quantum Institute. Lithium niobate devices were fabricated in the Center for Nanoscale Systems (CNS) at Harvard, a member of the National Nanotechnology Infrastructure Network, supported by the NSF.